\documentclass[fleqn,usenatbib]{mnras}
\usepackage{newtxtext,newtxmath}
\usepackage[utf8]{inputenc}
\usepackage{float}
\usepackage{graphicx}
\usepackage{amsmath}
\usepackage{datetime}
\usepackage[normalem]{ulem}
\usepackage{times}
\usepackage[export]{adjustbox}
\usepackage{soul}
\usepackage{caption}
\usepackage{dirtytalk}
\usepackage{gensymb}

\title[Electron Scattering in Centrifugal Magnetospheres]{Electron Scattering Emission in the Light Curves of Stars with Centrifugal Magnetospheres}

\author[I. Berry et al.]{I. D. Berry,$^{1}$\thanks{email: ianbass@udel.edu} 
S. P. Owocki,$^{1,2}$ 
M. E. Shultz,$^1$ 
A. ud-Doula$^3$
}

\date{%
    $^1$\textit{Department of Physics \& Astronomy, University of Delaware, Newark, DE 19716, USA}\\%
    $^2$\textit{Bartol Research Institute, University of Delaware, Newark, DE 19716, USA}\\%
    $^3$\textit{Penn State Scranton, 120 Ridge View Dr., Dunmore, PA 18512, USA}\\[2ex]%
}

\begin{document}

\maketitle

\begin{abstract}
    Strongly magnetic, rapidly rotating B-type stars with relatively weak winds form centrifugal magnetospheres (CMs), as the stellar wind becomes magnetically confined above the Kepler co-rotation radius. Approximating the magnetic field as a dipole tilted by an angle $\beta$ with respect to the rotation axis, the CM plasma is concentrated in clouds at and above the Kepler radius along the intersection of the rotational and magnetic equatorial planes. Stellar rotation can bring such clouds in front of the stellar disk, leading to absorption of order 0.1 magnitude ($\sim 10 \%$ of continuum flux). However some stars with prominent CMs, such as $\sigma$ Ori E, show an emission bump in addition to absorption dips, which has been so far unexplained. We show that emission can occur from electron scattering toward the observer when CM clouds are projected off the stellar limb. Using the Rigidly Rotating Magnetosphere model, modified with a centrifugal breakout density scaling, we present a model grid of photometric light curves spanning parameter space in observer inclination angle $i$, magnetic obliquity angle $\beta$, critical rotation fraction $W$, and optical depth at the Kepler radius $\tau_{\text{K}}$. We show that $\tau_{\text{K}}$ of order unity can produce emission bumps of the magnitude $\sim 0.05$ seen in $\sigma$ Ori E. We discuss the implications for modeling the light curves of CM stars, as well as future work for applying the radiative transfer model developed here to 3D MHD simulations of CMs.
\end{abstract}

\begin{keywords}
stars: early-type $-$ stars: massive $-$ stars: winds $-$ stars: magnetic field $-$ stars: chemically peculiar
\end{keywords}

\today

\section{Introduction} \label{Intro}
Hot, massive, and luminous stars of spectral type O and B have powerful and dense radiation-driven stellar winds \citep[e.g.][]{1975ApJ...195..157C,1995ApJ...454..410G,vink2001}. A small but significant fraction of these stars ($\sim$10\%) have magnetic fields \citep{2017MNRAS.465.2432G, 2019MNRAS.483.3127S} with strengths on the order of 100s of G to 10s of kG \citep{2019MNRAS.490..274S,2019MNRAS.483.3127S,2019MNRAS.489.5669P}. These magnetic fields largely exhibit simple topologies, with the majority being well approximated by dipoles tilted with respect to the rotational axis \citep{2019A&A...621A..47K}. The combination of a strong magnetic field and an ionized stellar wind allows plasma to become trapped within the closed magnetic loops leading to the formation of a circumstellar magnetosphere \citep{1987pbes.coll...82B,1997A&A...323..121B,2002ApJ...576..413U} which is forced to co-rotate with the photospheric magnetic field.\par 

The nature of such magnetospheres  depends on the rate of stellar rotation. Magnetic spectral type O stars tend to be slow rotators because the magnetic torque applied to their powerful stellar winds leads to rapid angular momentum loss \citep{2009MNRAS.392.1022U,2013MNRAS.429..398P}. The slow rotation of O stars means that their Kepler co-rotation radii ($R_\text{K}$), the distance at which the star's orbital and rotational period are equal, is typically well beyond the Alfv\'en radius ($R_{\rm A}$), the maximum distance of closed magnetic field loops. Stellar wind material trapped in such closed  loops has no centrifugal support and so falls back to the surface on dynamical timescales due to gravity, leading to a \say{dynamical magnetosphere} \citep[DM;][]{2002ApJ...576..413U,2013MNRAS.429..398P}.\par

 The less luminous B-type stars have smaller mass-loss rates and significantly longer spin down times than O stars. If their rotation is fast enough, i.e. a significant fraction of the orbital velocity at the stellar surface, then the Kepler co-rotation radius can be less than the Alfv\'en radius \citep{2008MNRAS.385...97U}. Material fed into the magnetosphere at or above the Kepler radius becomes trapped as centrifugal support stops it from falling back to the star, while the magnetic field stops it from flowing out. This trapped material  accumulates into a dense cloud in the region between
 $R_{\rm K}$ and $R_{\rm A}$, forming a
 \say{centrifugal magnetosphere} \citep[CM;][]{2005MNRAS.357..251T,2008MNRAS.385...97U,2013MNRAS.429..398P}.\par

The transition between DMs vs.\ CMs was first clearly illustrated in the MHD simulation parameter study by \citet[][see their Figure 9]{2008MNRAS.385...97U} for the 2D axi-symmetric case of a rotation aligned dipole.
This provided a dynamical confirmation of many of the features predicted by the analytic \emph{rigidly rotating magnetosphere} model (RRM) developed by \citet[][henceforth TO05]{2005MNRAS.357..251T},
which solved for the accumulation surface of a CM under the key simplifying assumption that the circumstellar magnetic field remains completely rigid to arbitrarily large radii (much as if $R_{\rm A} \rightarrow \infty$).\par

The B2\,Vp star $\sigma$ Orionis E (henceforth $\sigma$ Ori E) is the prototypical example of a star displaying photometric variability due to the presence of a circumstellar magnetosphere \citep{1978ApJ...224L...5L} forced into co-rotation by $\sigma$ Ori E's strong \citep[$\sim$10 kG;][]{2012MNRAS.419..959O} magnetic field. The photometric and magnetic periods are identical \citep{1978ApJ...224L...5L}. The light curve of $\sigma$ Ori E shows two major dips in brightness during its 1.19 day period, with light curve minima coinciding with magnetic nulls, i.e.\ rotational phases corresponding to a view along the magnetic equator. \cite{1978ApJ...224L...5L} suggested that all of these properties could be explained by corotating, magnetically trapped plasma, with absorption occuring when plasma clouds in the magnetic equatorial plane pass in front of the stellar disk. This understanding was demonstrated to be fundamentally correct by \cite{2005ApJ...630L..81T}, who successfully modeled the light curve of $\sigma$ Ori E obtained by \cite{1977ApJ...216L..31H} using the RRM model.\par

\cite{2008MNRAS.389..559T} used the RRM model to explore photometric modulation due to eclipsing for a comprehensive range of tilt angles, inclinations and rotation periods, demonstrating that eclipses can only be detected in stars with highly tilted magnetic fields seen with their rotational poles at large angles relative to the line of sight. Magnetospheric eclipses have also been studied by \cite{2020MNRAS.492.1199M}, who conducted a similar study of magnetospheric light curve modulation around magnetic O-type stars using the Analytical Dynamical Magnetosphere model developed by \cite{2016MNRAS.462.3830O}.\par

Eclipses from circumstellar magnetospheres are not the only, or even the most common, mechanism generating rotationally modulated photometric variability for magnetic stars. Such stars are invariably chemically peculiar \citep[CP; e.g.][]{2006A&A...450..763K,2019MNRAS.483.2300S}. $\sigma$ Ori E, which is a helium-rich star \citep[e.g.][]{2015MNRAS.451.2015O}, is no exception. Chemical peculiarities in magnetic stars tend not to be homogeneous across the stellar surface, instead they are clumped together in surface abundance patches. These surface abundance patches lead to photometric variation on the rotational timescale \citep[e.g.][]{2001A&A...378..113R,2019MNRAS.487.4695S}. Chemical abundance patches can be reproduced via surface chemical abundance maps obtained via Doppler imaging \citep{2017A&A...597A..58K}. \cite{2007A&A...470.1089K} showed that the surface distribution of silicon and helium was the cause of photometric variability of the CP star HD\,37776. The connection between chemical spots and and photometric variability has since been demonstrated for a number of other mCP stars \citep[e.g.][]{2009A&A...499..567K,2010A&A...524A..66S,2013A&A...556A..18K,2015A&A...584A..17P}. 

Further analysis of $\sigma$ Ori E's magnetosphere was done by \cite{2000A&A...363..585R}, with work recently done by \cite{2015MNRAS.451.2015O} to better fit the light curve by taking abundance patches into account. High resolution spectropolarimetry and magnetic Doppler imaging (MDI) \citep{2002A&A...381..736P} was used to map the star's surface magnetic field configuration. This \say{arbitrary} RRM accounted for departures from a purely dipolar geometry, which were reconstructed self-consistently with surface chemical abundance distributions via inversion of polarized and unpolarized line profile variations. A revised version of RRM was used which accepts arbitrary magnetic field configurations, referred to as the \emph{arbitrary-field rigidly rotating magnetosphere model} (ARRM). However, the ARRM model was unable to produce any significant improvements in the agreement between the modeled and observed light curves of $\sigma$ Ori E. Furthermore, surface abundance inhomogeneities were unable to make up for any discrepancies between the modeled and observed photometry. \par

The most significant discrepancy between the observed and modeled light curves for $\sigma$ Ori E is a small bump in brightness (extra $\sim$5\%) seen at phase $\sim$0.6 in the light curve, which is unaccounted for by any RRM model or surface abundance map. One obvious explanation may be that only absorption by the circumstellar magnetosphere was modeled by TO05 and \cite{2015MNRAS.451.2015O}.\par

Material enters magnetospheres via line-driven stellar winds along magnetic field lines into potential minima (TO05). How material exits the magnetosphere was first investigated by \cite{1984A&A...138..421H} who conjectured that magnetic field loops break open from an excess of plasma. While RRM assumes the magnetic field remains perfectly rigid regardless of the amount of material built up, in practice all magnetic fields have finite strength so the growing centrifugal force from accumulating material must at some point overwhelm the magnetic tension. Known as centrifugal breakout (CBO), a first quantitative analysis was included in the Appendix of TO05, which provided expressions for the breakout density, timescale, as well as the breakout-limited asymptotic mass. CBO was found to occur on similar timescales in MHD simulations by \citet{2008MNRAS.385...97U,2006ApJ...640L.191U}, who further predicted X-ray emission when magnetic field loops snap back together after breakout occurs. However, whereas TO05 assumed a complete emptying of material beyond the Kepler radius, \cite{2008MNRAS.385...97U} found that material can fall back to the surface of the star after breakout via magnetic recombination, which can limit the overall mass of the magnetosphere.\par

The notion of large-scale emptying by CBO events was challenged by \cite{Townsend_2013} who used observations from the \emph{MOST} \citep[Microvariability and Oscillations in Stars;][]{2003PASP..115.1023W} satellite to measure the photometric variability of $\sigma$ Ori E on a time-scale of three weeks. This found no evidence of the catastrophic magnetospheric reorganization suggested by the 2D MHD simulations explored by \cite{2006ApJ...640L.191U} and \cite{2008MNRAS.385...97U}.
\citeauthor{Townsend_2013} further argued that the absorption depth in the light curve of $\sigma$ Ori E requires a magnetospheric mass much less than that associated with the asymptotic mass from breakout. This led to the development of a different model for CM plasma transport by \citet{2018MNRAS.474.3090O} involving gradual `leakage' via diffusion and drift across magnetic field loops.
However, a recent follow-up analysis by \citet{2020MNRAS.499.5366O} showed that a combination of factors led \citet{Townsend_2013} to overestimate the CBO mass and density, and that these were in fact consistent with existing empirical constraints.\par

Building upon the extensive empirical analysis of H$\alpha$ emission from a large sample of magnetic B-stars conducted by \citet{2020MNRAS.499.5379S}, \citet{2020MNRAS.499.5366O} also showed that the CBO model can explain how both the onset and strength of H$\alpha$ emission scale with stellar, magnetic, and rotational parameters, most notably the field strength at the Kepler radius, 
$B_{\text{K}}$. \citet{2020MNRAS.499.5379S} further showed
that H$\alpha$ emission is relatively insensitive to effective temperature, luminosity and mass-loss rate, contrary to what is expected from the leakage models explored by \citet{2018MNRAS.474.3090O} \citep[see also the exploration of the leakage question by][]{2013MNRAS.429..398P}.
These results strongly favor centrifugal breakout regulating plasma transport in the centrifugal magnetosphere. However, instead of large-scaled sporadic emptying events suggested by initial 2D models, both empirical analysis \citep{2020MNRAS.499.5379S} and theoretical arguments \citep{2020MNRAS.499.5366O} indicate that CBO is a quasi-steady process, characterized by frequent, small-scale outer-region ejections spread through the magnetosphere, which is maintained more or less constantly at the breakout density. \par

This CBO model suggests a density distribution that differs significantly from what was utilized in the original RRM analysis by TO05. Specifically, TO05  assumed that the magnetosphere was emptied out at some arbitrary time in the past, with the local density at any later time set by the local wind feeding rate, which varies in proportion to the local field strength. By comparison, CBO predicts a radial surface density that scales with the \emph{square} of the magnetic field strength. This gives a steeper radial decline, but an overall higher density in the CM \citep{2020MNRAS.499.5366O}. With a higher density comes the possibility that CMs can be optically thick in the continuum, even with the relatively low wind feeding rate of B-type stars.

This provides a central motivation to explore the emission from CMs via free-free electron scattering from dense material projected off the limb, while also still accounting for the net absorption from material projected across the stellar disk.
We follow a similar strategy to \cite{2008MNRAS.389..559T,2020MNRAS.492.1199M}, generating photometric light curves in a thorough parameter study varying inclination angles, magnetic obliquities, optical depths and rotation periods. 
Section \ref{Methods} explains how we've modeled magnetospheres, how we calculate appropriate optical depths in the continuum, and reviews our radiative transfer treatment with a scattering source function. Section \ref{aligned} presents results for the simple case of a field-aligned dipole. Section \ref{tilted} presents the light curve parameter study for tilted dipoles. This is followed by a general discussion in Section \ref{disc} and a summary in Section \ref{sum}.

\section{Magnetospheres and Radiative Transfer} \label{Methods}
\subsection{Rigidly Rotating Magnetosphere with modified density}\label{RRMCBO}

The underlying CM models for our radiative transfer calculations of electron scattering emission are grounded in the Rigidly Rotating Magnetosphere (RRM) formalism developed by TO05.
For a star with mass $M_\ast$, equatorial radius $R_\ast$, and rotation frequency $\Omega$, we can compare the surface equatorial rotation speed $V_{\rm rot} \equiv \Omega R_\ast$ and the near-surface orbital speed $V_{\rm orb} \equiv \sqrt{GM_\ast/R_\ast}$, defining a critical rotation fraction and associated Kepler co-rotation radius,
\begin{equation}
W \equiv \frac{V_{\rm rot}}{V_{\rm orb}} ~~ ; ~~ 
R_{\rm K} = W^{-2/3} R_\ast
\, .
\end{equation}

Above $R_{\rm K}$ the outward centrifugal acceleration from rigid-body co-rotation will exceed the local inward gravitational acceleration.
Thus, for rigid magnetic loops that extend above this Kepler radius, the RRM model envisions that stellar wind material driven from the loop footpoints will accumulate into a hydrostatic stratification about minima of the effective rotational+centrifugal potential along that field line.

In the original RRM implementation by TO05,
the relative density distribution about the resulting potential minimum surface is assumed to scale with the local feeding rate by the stellar wind, much as if the material had been accumulating for an arbitrary fixed time after some previous complete emptying;
loops without such a potential minimum have no accumulation and so are assigned zero (or negligible) density.

\begin{figure}
    \centering
    \includegraphics[width = \columnwidth]{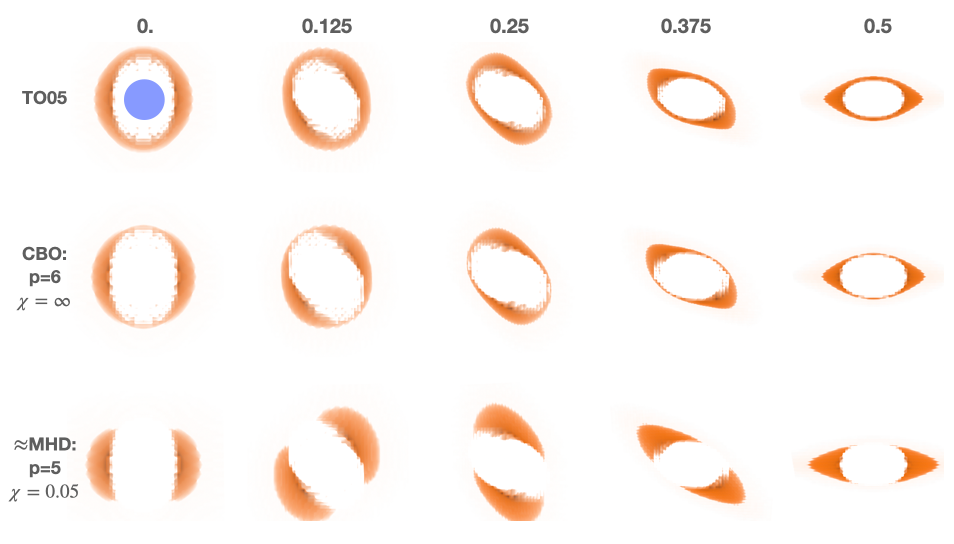}
    \caption{For magnetic obliquity $\beta = 60 \degree$, plots of integrated column
    density for a CM viewed from inclination $i=60 \degree$ at various rotational phases up to 0.5 (after which they simply reverse), 
    as labeled at the top of each column.
The rows compare the density scaling assumed in the original RRM analysis by TO05
(top) with one using a radial power index $p=6$ to mimic the $\rho \sim B^2 \sim r^{-6}$ scaling inferred from the CBO analysis by \citet{2020MNRAS.499.5366O} (middle), along with the model used here with $p=5$ and $\chi=0.05$, which mimics the radial and aziumthal density variation found from 3D MHD simulations (bottom). In all models, the innermost non-zero density occurs at the Kepler corotation radius $R_{\rm K} = W^{-2/3} R_\ast$. To illustrate the relative scales, the upper left panel includes the stellar disk for a critical rotation fraction $W=1/2$.
    }
    \label{TO05vCBO}
\end{figure}

For the simple example case of a dipole field with magnetic axis tilted by an angle $\beta=60 \degree$ from the rotation axis, the top row of Figure \ref{TO05vCBO} illustrates the associated TO05 variation of column density with rotational phase, when viewed from an inclination angle $i=60 \degree$.
For the view down the magnetic axis at phase $\phi =0$,  the column density distribution takes an oblong form, with the narrow, innermost CM edge along the horizontal axis formed from the intersection of the magnetic and rotational equators; 
along the vertical axis, the CM edge is further from the center, with a lower density that reflects the weaker feeding from the lower wind mass flux, which for rigid-field outflow scales with the local field strength $\rho \sim B \sim r^{-3}$.
As also illustrated in the corresponding Figure 4 in TO05, the overall effect is that material is largely concentrated in ``clouds'' on each side of the common rotational and magnetic equatorial axis.

Building on scalings derived in the appendices of TO05, \citet{2020MNRAS.499.5366O} has showed that both the onset and strength of Balmer-$\alpha$ line emission can be best explained by a centrifugal breakout (CBO) scaling, for which the local accumulated density varies with the {\em square} of the local field strength, giving now for a dipole $\rho \sim B^2 \sim r^{-6}$.
For this simple CBO scaling, the middle panel of Figure \ref{TO05vCBO} compares the column density distributions vs. rotational phase for the same tilted dipole field with $\beta= 60 \degree = i$.
Note that the steeper decline in density further exaggerates the oblong distribution of column density, now with an even greater concentration of density into equatorial clouds.

Recent 3D MHD simulations \citep{2021mobs.confE..33U}, indicate an even stronger concentration of density into clouds along the common equatorial axis. 
The potential minima along this axis occur at the loop apex, at magnetic co-latitude $\theta_o = 90 \degree$, with thus the magnetic field orthogonal to the direction of the local centrifugal acceleration, providing the most effective confinement.
Away from this common equatorial axis, and most particularly for directions orthogonal to it, the potential minimum is displaced away from the loop apex, with $|\theta_o - 90 \degree| > 0$; the smaller angle between field and centrifugal acceleration makes for less effective confinement, and so greater centrifugal escape.

To account for this, we have developed the following parameterized fitting form for the surface density on a local potential minimum with radius $r$ and magnetic co-latitude $\theta_o$,
\begin{equation}
\sigma (r, \theta_o) = \sigma_{\rm K} \left ( \frac{r}{R_{\rm K}} \right )^{-p} \, \exp(-\cos^2 \theta_o/\chi ) 
\, .
\label{eq:sigrtho}
\end{equation}

Here $\sigma_{\rm K}$ is the associated peak surface density at the Kepler radius $r=R_{\rm K}$, for a loop lying in the rotational equator, for which the potential minimum occurs at the loop apex that also defines the magnetic equator, i.e. with $\theta_o = 90 \degree$ and thus $\cos \theta_o = 0$.
The density decline away from this apex falls as a gaussian in $\cos \theta_o$,  with the parameter $\chi$ representing a kind of ``latitudinal scale length";  the radial dropoff is now taken to follow a general power-law form with index $p$.
We find that the results of 3D MHD simulations are roughly fit by parameters $p=5$ and $\chi=0.05$, and so adopt this as our standard model for the electron scattering calculations below.
The bottom panel of Figure  \ref{TO05vCBO} compares the column density variations for this standard model aimed at mimicking MHD results.

Following TO05, in all 3 models the volume density $\rho$ along a field line is assumed to follow a hydrostatic stratification about the local potential minimum, as given by TO05 equation (25).
For gravitationally scaled potential difference $\Delta \Psi (s)$ at field line position $s$, we have
\begin{equation}
\rho(s) = \rho_{\rm m} \,  \exp  (-\Delta \Psi (s)/\epsilon )
\, ,
\label{eq:rhos}
\end{equation}
where for a constant temperature $T$ and molecular weight $\mu$, we have defined the ratio of thermal energy to gravitational escape energy at the Kepler radius,
\begin{equation}
    \epsilon \equiv \frac{k T R_{\rm K}}{G M_* \mu}~,
    \label{eq:epsilon}
\end{equation}
with $k$ and $G$ the Boltzmann and gravitational constants, respectively. In direct analogy with the transformation between TO05 equations (25) to (27), equation (\ref{eq:rhos}) here leads to a gaussian stratification about the local minimum potential; the energy ratio $\epsilon$ sets the disk thickness, with  scale height $h_{\rm m}$ given by TO05 equations  (28) and (29).  
Our implementation here sets $\epsilon = 0.01$, giving then $h_{\rm m}/R_{\rm K} \approx \sqrt{2\epsilon/3} = 0.082$. 
In terms of the local surface density $\sigma (r,\theta_o)$ from equation (\ref{eq:sigrtho}). The stratification of volume density a locally normal distance $z$ away from the potential minimum surface is thus given by:
\begin{equation}
\rho(z) = 
\frac{\sigma (r,\theta_o)}{\sqrt{\pi} h_{\rm m}} \, \exp [-(z/h_{\rm m})^2]
\, .
\label{eq:rhoz}
\end{equation}

\noindent Integration over all $z$ then recovers this local surface density $\sigma(r,\theta_{\rm o})$.

\subsection{Optical Depth of the Centrifugal Magnetosphere} \label{escattering}
As we quantify from detailed models in section \ref{tilted}, achieving a non-negligible emission from electron scattering requires a CM that is at least marginally optically thick to electron scattering opacity\footnote{For fully ionized H and He, $\kappa_{\rm e} = 0.2 (1+X)  = 0.344$\,cm$^2$/g for standard hydrogen mass fraction $X=0.72$.} $\kappa_{\rm e}$.
Let us thus here consider what the overall CBO scalings for CM density imply for the electron scattering optical depth, given parameters relevant for strongly magnetic B-stars like $\sigma$~Ori~E.

As noted by \citet{2020MNRAS.499.5366O}, for the simple case of an aligned dipole ($\beta=0$) the appendices of TO05 provide a basic CBO scaling for the peak surface density at the Kepler radius in terms of the equatorial magnetic field strength $B_{\rm K} = B_{\rm eq}/R_{\rm K}^3$ and stellar gravity $g_{\rm K}$ at this distance,
\begin{equation} \label{sigstar}
    \sigma_{\rm K} = c_{\rm f} \, \frac{B_{\text{K}}^2}{4\pi g_{\text{K}}}\text{,}
\end{equation}
wherein $c_{\rm f}$ is an order-unity correction factor, calibrated by comparison with MHD simulations (by which \citet{2020MNRAS.499.5366O} find $c_{\rm f} \approx 0.3$).
We can write the associated electron scattering optical depth in terms of the gravity $g_\ast = GM_\ast/R_\ast^2$ and polar field strength $B_\ast = 2 B_{\rm eq,\ast}$ at the stellar surface radius $R_\ast$,
\begin{equation}
\tau_{\rm K} \equiv \tau_e(R_{\rm K} ) 
= c_{\rm f} \kappa_{\rm e} \, \frac{B_\ast^2}{16 \pi g_\ast} \, \left ( \frac{R_{\rm K}}{R_\ast} \right )^{-4}
= 0.11 \,  \left ( \frac{c_{\rm f}}{0.3} \right ) \, \left (  \frac{B_\ast}{{\rm kG}} \right )^2 W^{8/3}
\, ,
\label{eq:tauk}
\end{equation}
where the last equality applies for typical  parameters $M_\ast = 8 {\rm M_\odot}$ and $R_\ast = 3.5 {\rm R_\odot}$ for a B2\,V star.

\begin{figure}
    \centering
    \includegraphics[width = \columnwidth]{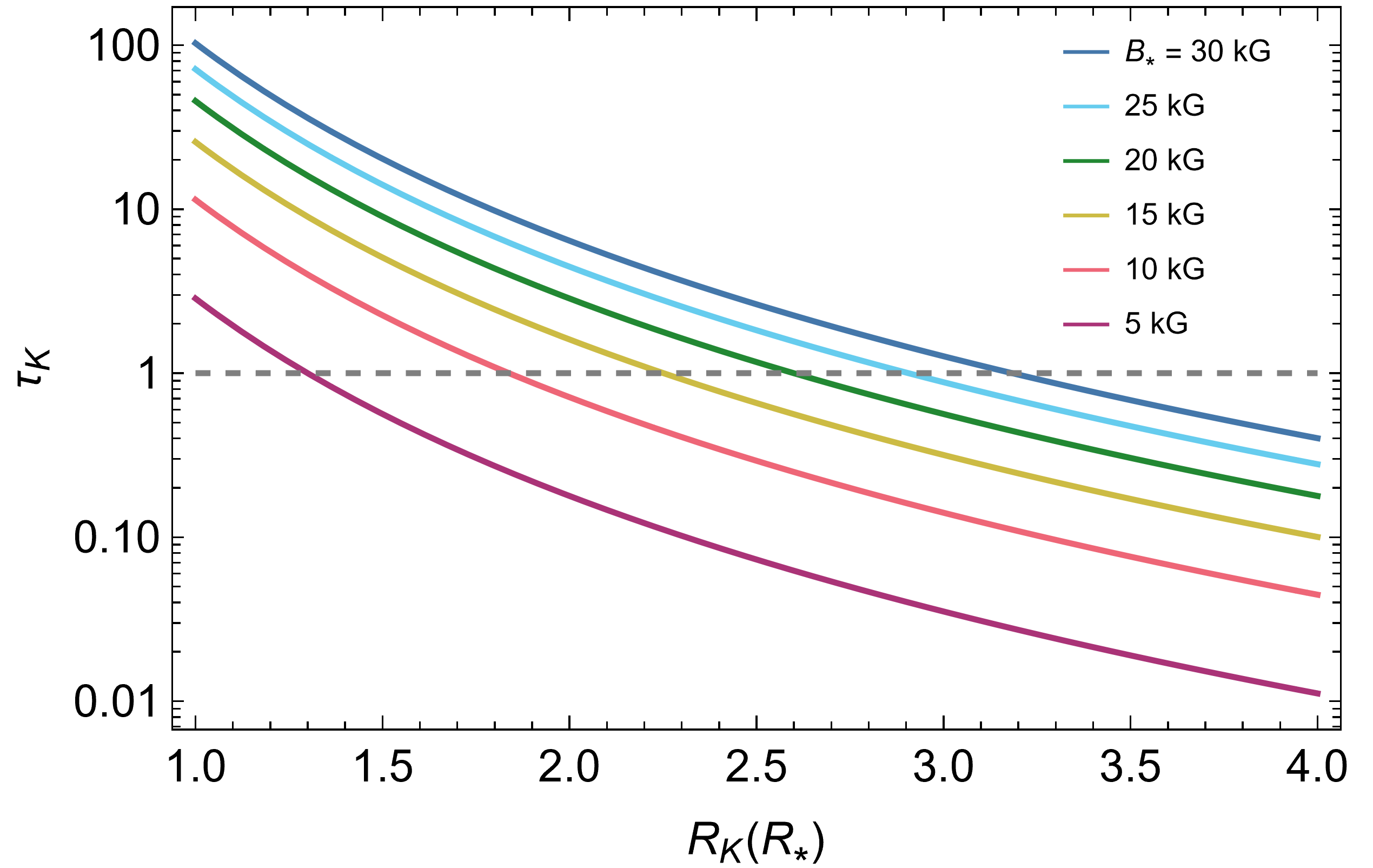}
    \caption{ $\tau_{\text{K}}$ vs. $R_{\text{K}}$ in units of $R_*$ at varying surface magnetic polar field strengths, and with the parameters of a typical B2\,V star ($\text{M}_* = 8~\text{M}_{\sun}$ and $\text{R}_* = 3.5~\text{R}_{\sun}$). The dashed gray line denotes $\tau_{\rm K} = 1$. $\tau_{\rm K}$ will be of order unity for common combinations of $W$ and $B_*$ (e.g. $W = 0.5$ and $B_* = 10$~kG, which gives us the justification to model CMs with optical depths $\tau_{\text{K}} = 0.5, 1$ and 2. It is possible for centrifugal magnetospheres to be significantly optically thick ($\tau_{\rm K} \sim 100 $) in the continuum due to both an extremely strong magnetic field ($\sim 30$ kG) and critical rotation, however such stars have not yet been observed.}
    \label{taukVar}
\end{figure}

Figure \ref{taukVar} plots the associated variation of $\tau_{\rm K}$ vs. $R_{\rm K}/R_\ast$ for various polar field strengths up to the maximum inferred $B_\ast \approx 30$~kG. For a  star with such extreme magnetic fields and near-critical rotation, this indicates the CM could in principle become quite optically thick to electron scattering, with $\tau_{\rm K} \rightarrow 100$. 
But such extreme combinations are unlikely \citep[the most rapidly rotating CM host stars yet found have $W \sim 0.5$ and $B_{*} \sim 10$~kG;][]{2019MNRAS.490..274S}, so for a more common range of parameters, CMs are expected to be at most only marginally optically thick to electron scattering, i.e. $\tau_{\rm K} \gtrsim 1$.

Equation (\ref{eq:tauk}) can be readily solved for the polar surface field in terms of the critical rotation fraction $W$ and stellar parameters relative to their solar values,
\begin{equation} \label{BtauW} 
B_\ast = {\cal B} \, 
\frac{\sqrt{(M_\ast/M_\odot) \tau_{\rm K}/c_{\rm f}}}{(R_\ast/R_\odot) W^{4/3}}
= 2950 \, {\rm G} \, \sqrt{\frac{0.3\, \tau_{\rm K}}{c_{\rm f}}} \, W^{-4/3} 
\, ,
 \end{equation}
 where the latter equality again assumes the above typical B2\,V stellar parameters.
  The constant,
 \begin{equation}
     {\cal B}  \equiv \sqrt{\frac{16\pi g_{\text{\sun}}}{\kappa_{\text{e}}}} 
     \approx 2~\text{kG}~,
  \end{equation}
provides a canonical field strength in terms of the sun's typical surface gravity.
Equation (\ref{BtauW}) shows that, even for near-critical rotation $W \sim 1$,  making a CM optically thick in electron scattering generally requires kG fields, with the required field increasing as $1/W^{4/3}$ for subcritical rotation $W<1$.

\begin{figure}
     \centering
     \includegraphics[width = \columnwidth]{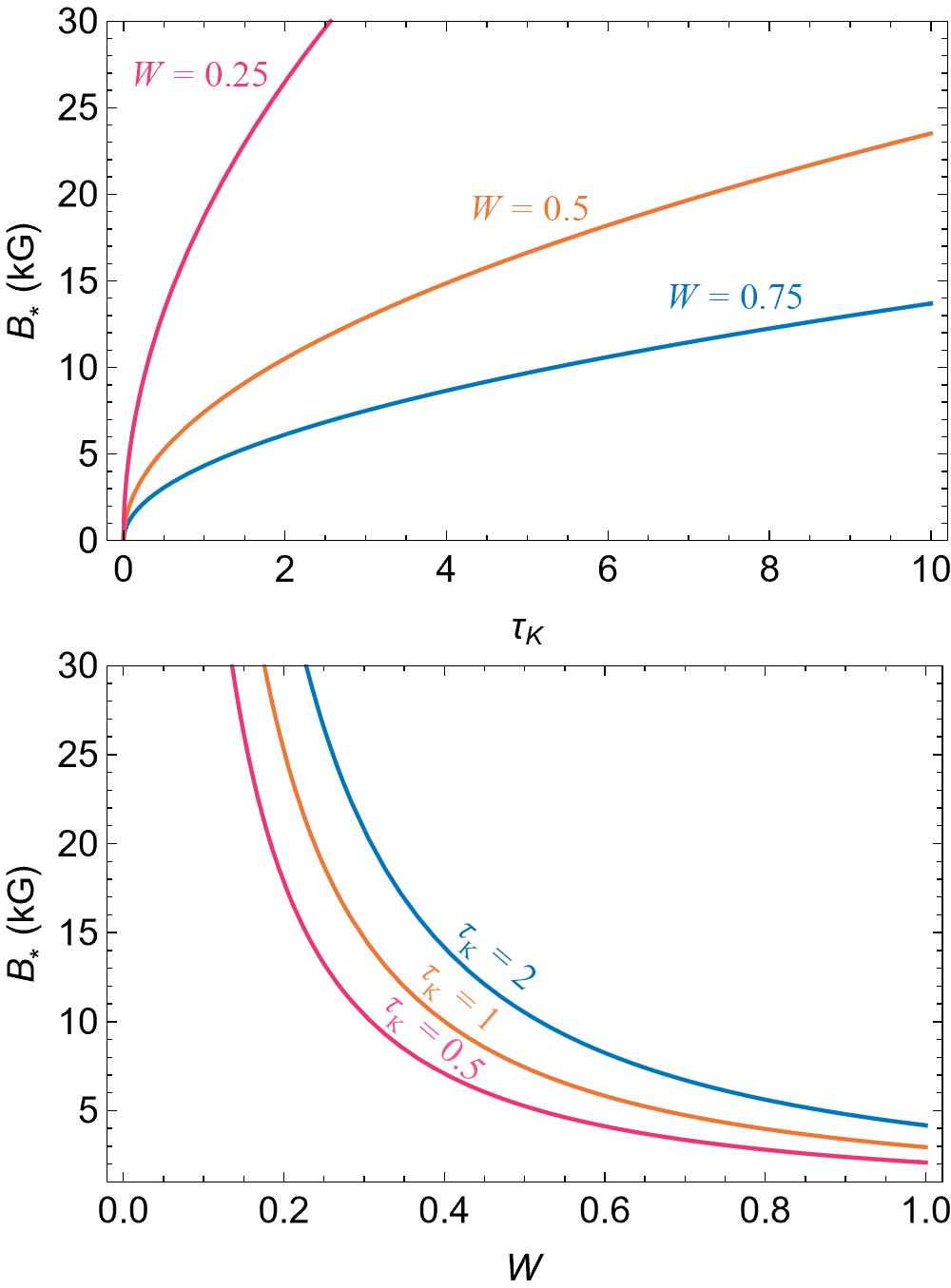}
     \caption{\emph{Top}: Surface polar field strength $B_*$ vs. $\tau_{\text{K}}$ with $W = 0.25$, 0.5 and 0.75. The stellar parameters of a typical B2\,V star are the same as in Figure \ref{taukVar}. $B_*$ increases monotonically with $\tau_{\text{K}}$, however the steepness of the increase is significantly dependent on $W$. \emph{Bottom}: $B_*$ as a function of $W$ with $\tau_{\text{K}} = 0.5$, 1 and 2. We are limited to a maximum $W = 1$, meaning a minimum $B_*$ is required for a given $\tau_{\text{K}}$. Likewise, with a maximum $B_*$ = 30 kG, a minimum $W$ is needed for a given $\tau_{\text{K}}$.}
     \label{tauBVar}
 \end{figure}

To illustrate the overall values and trends, for our canonical B2V star parameters Figure \ref{tauBVar} plots $B_\ast$ as a function of $\tau_{\rm K}$ (upper panel) or $W$ (lower panel), for select fixed values of the other parameter. Generally, $W$ plays a significant role in determining how large $B_*$ needs to be in order for a CM to be optically thick ($\tau_{\rm K} \geq 1$). For relatively slow rotation (e.g. $W = 0.25$) we see that the minimum $B_*$ for an optically thick CM is $\approx 15$ kG, while for faster rotation (close to critical), the minimum field strength is a few kG. The bottom panel of Figure \ref{tauBVar} shows similar trends to those of the top panel, i.e. it's easier to have an optically thick CM with higher values of $B_*$ and $W$.
  
\subsection{Radiative Transfer for Electron Scattering}

Let us next review the radiative transfer methods we use to derive photometric light curves that account for electron scattering from our 3D CBO-modified RRM models of the density distribution in CMs.

For  observer inclination $i$ at each rotational phase of a given model, we define the density $\rho(x,y,z)$ on a Cartesian grid, with the observer at $z \rightarrow -\infty$, and the y-direction along the projected rotation axis.
For each $x,y$ position on the sky, we first compute the variation of optical depth along the line-of-sight,
\begin{equation}
\tau(x,y,z) = \int_{-\infty}^z \, \kappa_{\rm e}  \rho(x,y,z') dz'
\, .
\label{eq:tauxyz}
\end{equation}
The observed intensity is then computed from the formal solution of the equation of radiative transfer,
\begin{equation}
I(x,y) = I_\ast e^{-\tau(x,y,z_\ast)}  + \int_{-\infty}^{z_{\rm m}} S(x,y,z') \, e^{-\tau(x,y,z')} \, \kappa_{\rm e} \rho(x,y,z') \, dz'
\, ,
\label{eq:Ixy}
\end{equation}
where $z_\ast = - \sqrt{R_\ast^2 - x^2 -y^2}$ is the $z$ position of the stellar surface for rays $\sqrt{x^2+y^2} \le R_\ast$ intersecting the stellar disk, which is taken to have a fixed intensity $I_\ast$,  and $z_{\rm m} = z_\ast$ for disk-crossing rays, and $+\infty$ otherwise.

Previous pure-absorption models have effectively assumed no source function, $S = 0$. For pure scattering, this is given by the angle-averaged intensity, $S=J$, and since $J$ itself depends on $I$, a general approach requires a solution of coupled integro-differential equations, e.g. by $\Lambda$-iteration  \citep{hubeny_mihalas_2015}.
But for our models with only moderately optically thick CMs, and thus typically just a single CM scattering of  the radiation source from the stellar core, $J$ is well approximated by the diluted mean intensity from the core, giving
\begin{equation}
S \approx J_\ast (r) = I_\ast \frac{1- \mu_\ast(r)}{2} ~~ ; ~~ \mu_\ast (r) \equiv \sqrt{1 - R_\ast^2/r^2}
\, ,
\label{eq:Jast}
\end{equation}
where $\mu_\ast$ is the cosine of the angular radius of the stellar core at radius $r = \sqrt{x^2+y^2+z^2}$.

Finally, the observed net flux relative to that from the stellar core is obtained by the area integral of the intensity,
\begin{equation}
f = \frac{1}{\pi R_\ast^2 I_\ast} \, \int_{-\infty}^\infty dx \int_{-\infty}^\infty dy \, I(x,y)
\, .
\label{eq:f}
\end{equation}
In practice these integrals are carried out as simple Simpson-rule sums over a discrete grid with a uniform step size $\Delta x = \Delta y = 0.05 R_{\rm K}$.
Because of the steep radial decline of density in these CBO models ($\rho \sim r^{-5}$), we find it sufficient to limit the computation to $\pm 2 R_{\rm K}$ in each dimension. For the line of sight integration in the $z$-direction, tests against higher-order Runge-Kutta integrations indicate that, for our marginally optically thick CMs, grid steps $\Delta z = h/4$ set to resolve the disk scale height $h$ are adequate.

\section{Modulation of Absorption and Emission}
\subsection{Aligned-Dipole Case} \label{aligned}
\begin{figure}
    \centering
    \includegraphics[width = \columnwidth]{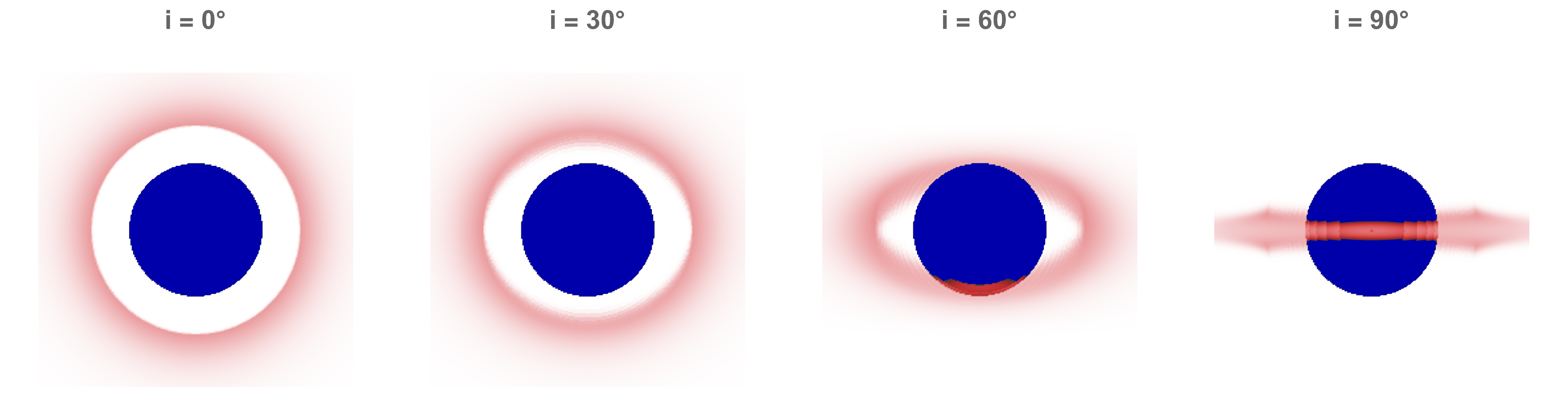}
    \caption{Surface intensity in the plane of the sky for inclination angles ranging from rotational pole-on ($i=0 \degree$) to along the equator ($i=90 \degree$), for a rotationally aligned dipole ($\beta=0^\circ$) with critical rotation fraction $W=0.5$, optical depth $\tau_{\rm K} = 2$ and disk thickness parameter $\epsilon = 0.5$. The blue disk representing the visible hemisphere of the star. The red ring indicates the CM out to $\pm 1.5 R_{\rm K}$. At low inclination, only emission occurs. When $i \geq 60^{\circ}$, absorption takes over as the magnetosphere eclipses the photosphere.}
    \label{fig:intInc}
\end{figure}

To illustrate the application of this scattering radiative transfer to our models of CMs, let us focus first on the relatively simple case of a rotationally aligned dipole ($\beta = 0 \degree$). In this case the disk has a 2D axisymmetry with no variation in azimuthal coordinate $\phi$, and thus no variation in rotational phase. However, scattering in the CM means that the observed flux will depend on the observer inclination angle to the rotational axis $i$. For the quoted choices of CM parameters, Figure \ref{fig:intInc} illustrates the surface brightness seen from various such observer inclinations.

\begin{figure}
    \centering
    \includegraphics[width = \columnwidth]{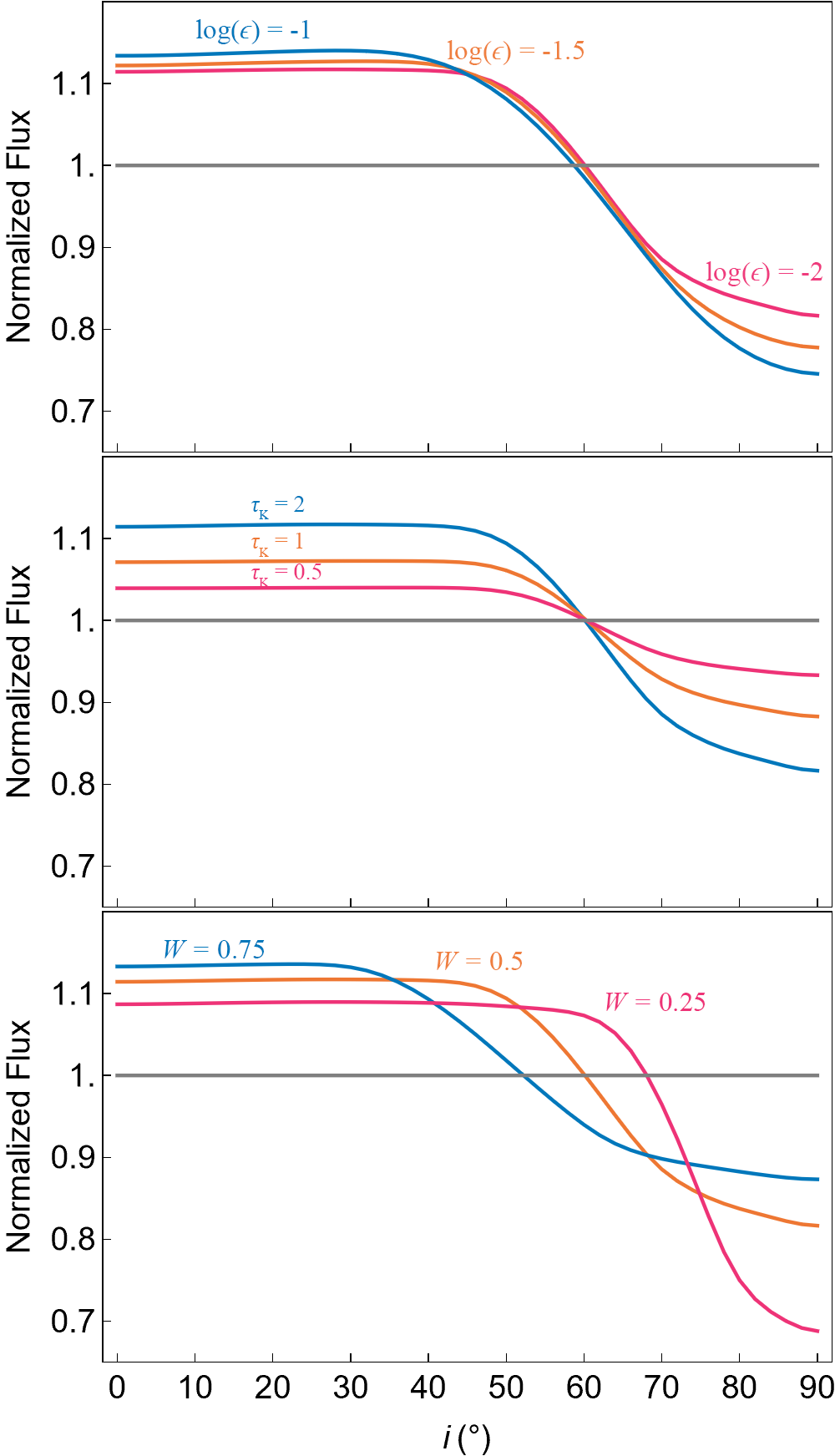}
    \caption{\emph{Top}: Flux normalized to the photospheric flux  vs. inclination angle $i$ for variation of $\epsilon$ with fixed $W = 0.5$ and $\tau_{\rm K} = 2$. The gray horizontal line denotes the constant stellar flux. Increasing $\epsilon$ slightly increases the amount of both emission and absorption. Furthermore absorption starts at a shallower $i$ at larger $\epsilon$ as the disk is physically thicker. \emph{Middle}: Same as top, except for $\tau_{\rm K}$ variation and fixed $\epsilon = 0.01$ and $W = 0.5$. Increasing $\tau_{\rm K}$ has the effect of increasing emission as well as absorption according to the solution to radiative transfer (equation \ref{eq:Ixy}). For small $\tau_{\rm K}$ variation, absorption begins at the same $i$. \emph{Bottom}: Same as top and middle except with $W$ variation and fixed $\tau_{\rm K} = 2$ and $\epsilon = 0.01$. Increasing $W$ has the effect of eclipses starting sooner as the distance between the stellar surface and the Kepler radius decreases. Emission increases, but absorption decreases as less of the star is eclipsed by the CM. Maximum emission in all panels is $\sim 10\%$ above the continuum.}
    \label{tauandW}
\end{figure}

 Figure \ref{tauandW} illustrates how the variation of flux vs. inclination is affected by the choice of these key CM parameters: Kepler optical depth $\tau_{\rm K}$, critical rotation fraction $W$, and disk thickness parameter $\epsilon$.
 For fixed $\tau_{\rm K} = 2$ and $W = 0.5$, the top panel of Figure \ref{tauandW} shows the variation of flux with inclination for 3 selected values of the disk thickness parameter. Note that higher thicknesses leads to stronger emission at low $i$ and deeper absorption at high $i$, but in all cases the overall trend shows a flux decline with increasing $i$.

For fixed $\epsilon = 0.01$ and $W = 0.5$, the middle panel of Figure \ref{tauandW} shows the variation of flux with $i$ for 3 selected values of $\tau_{\rm K}$. Note that higher values of $\tau_{\rm K}$ leads to higher emission at low $i$ and higher absorption at higher $i$, which is to be expected due to the solution to radiative transfer (\ref{eq:Ixy}). Also note that absorption starts roughly at the same time in phase for each $\tau_{\rm K}$. In all cases the overall trend shows a flux decline with increasing $i$.

For fixed $\tau_{\rm K} = 2$ and $\epsilon = 0.01$, the bottom panel of Figure \ref{tauandW} shows variation of flux with $i$ for 3 selected values of $W$. Note that higher values of $W$ leads to higher emission at low $i$ but lower absorption at higher $i$, both effects being due to the inner edge of the CM moving closer to the stellar surface at larger $W$. Note that in all cases the overall trend shows a flux decline with increasing $i$. 

\subsection{Tilted-Dipole Case} \label{tilted}

\begin{figure}
    \centering
    \includegraphics[width = \columnwidth]{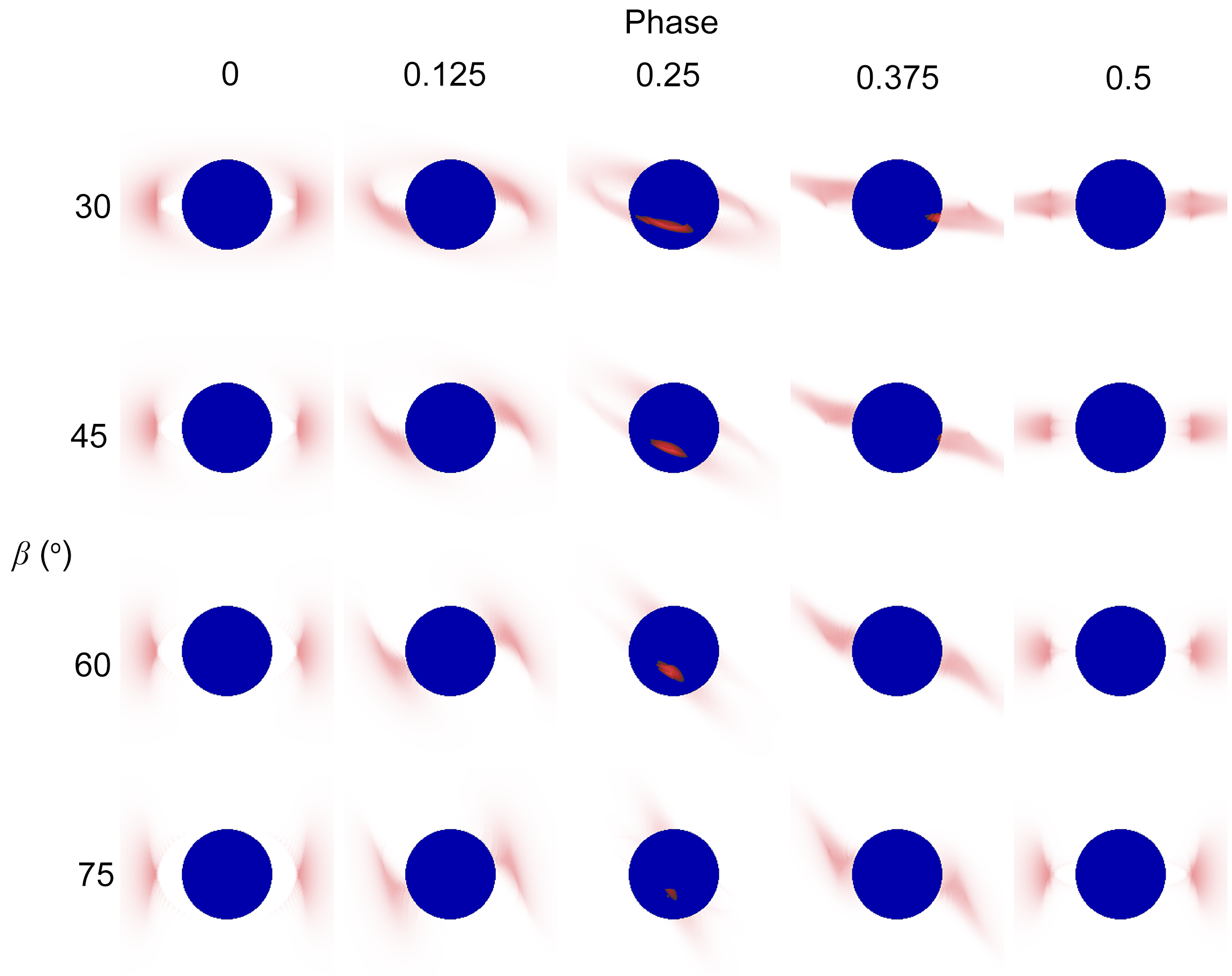}
    \caption{\emph{Top}: Surface intensity displayed at various $\beta$ and rotational phases, with fixed $i = 75^{\circ}$. The blue disk denotes the star, and red denotes the CM. Due to the lack of 2D axisymmetry for $\beta \neq 0$, periods of emission and absorption occur as the CM appears off and on the stellar disk respectively.}
    \label{surfaceint}
\end{figure}

\begin{figure*}
    \centering
    \includegraphics[width=\linewidth]{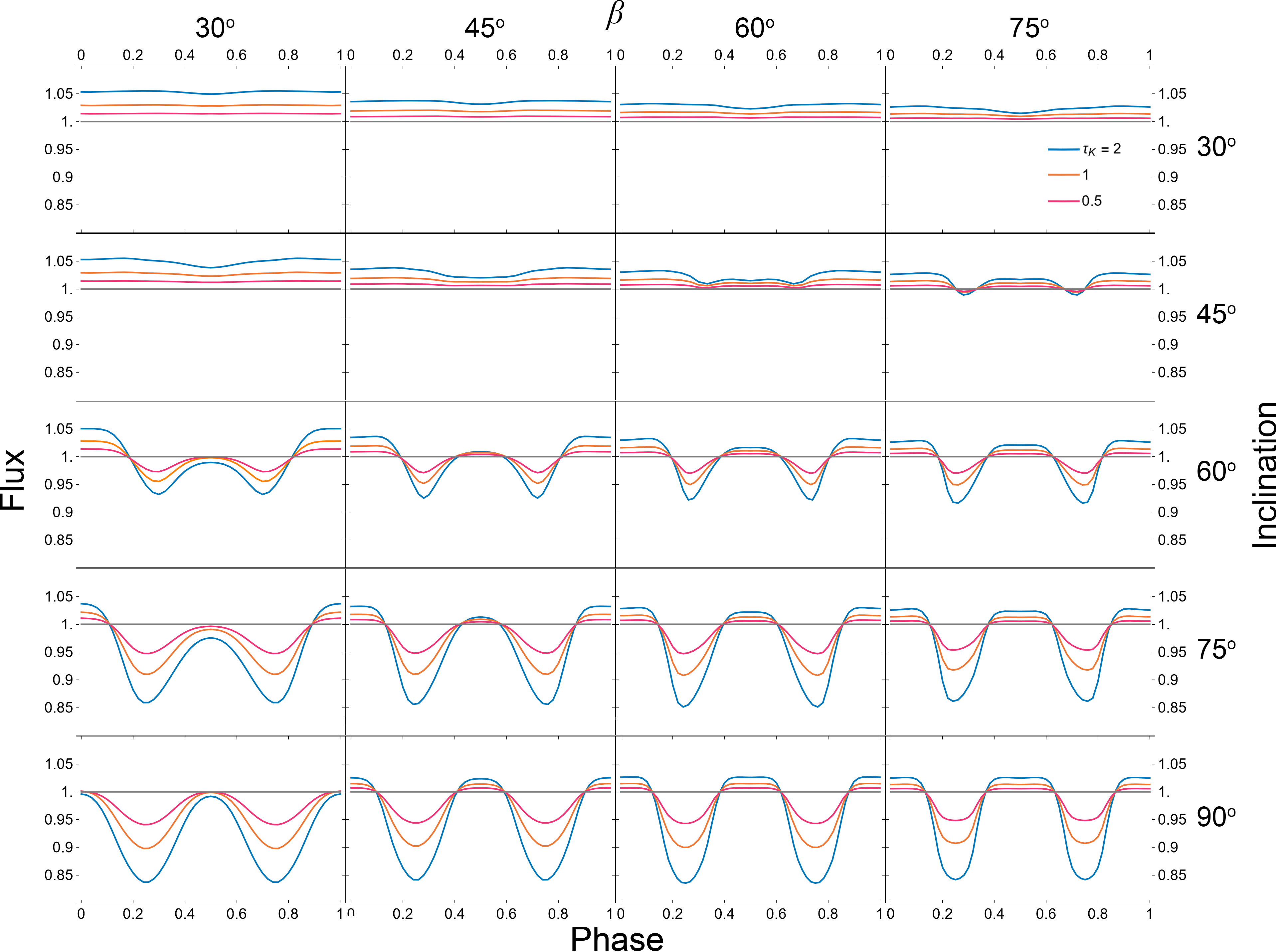}
    \caption{Light curves with $\tau_{\text{K}} = 0.5, 1$ and 2, with fixed $W = 0.5$. A horizontal gray line is placed at flux = $F_* \equiv 1$ to mark the transition from emission to absorption. Individual panels show models with $\beta = 30^{\circ}, 45^{\circ}, 60^{\circ}$ and $75^{\circ}$ as well as inclinations $i$ of the same angles (with the addition of $90^{\circ}$).}
    \label{lightcurvestau}
\end{figure*}

\begin{figure*}
    \centering
    \includegraphics[width=\linewidth]{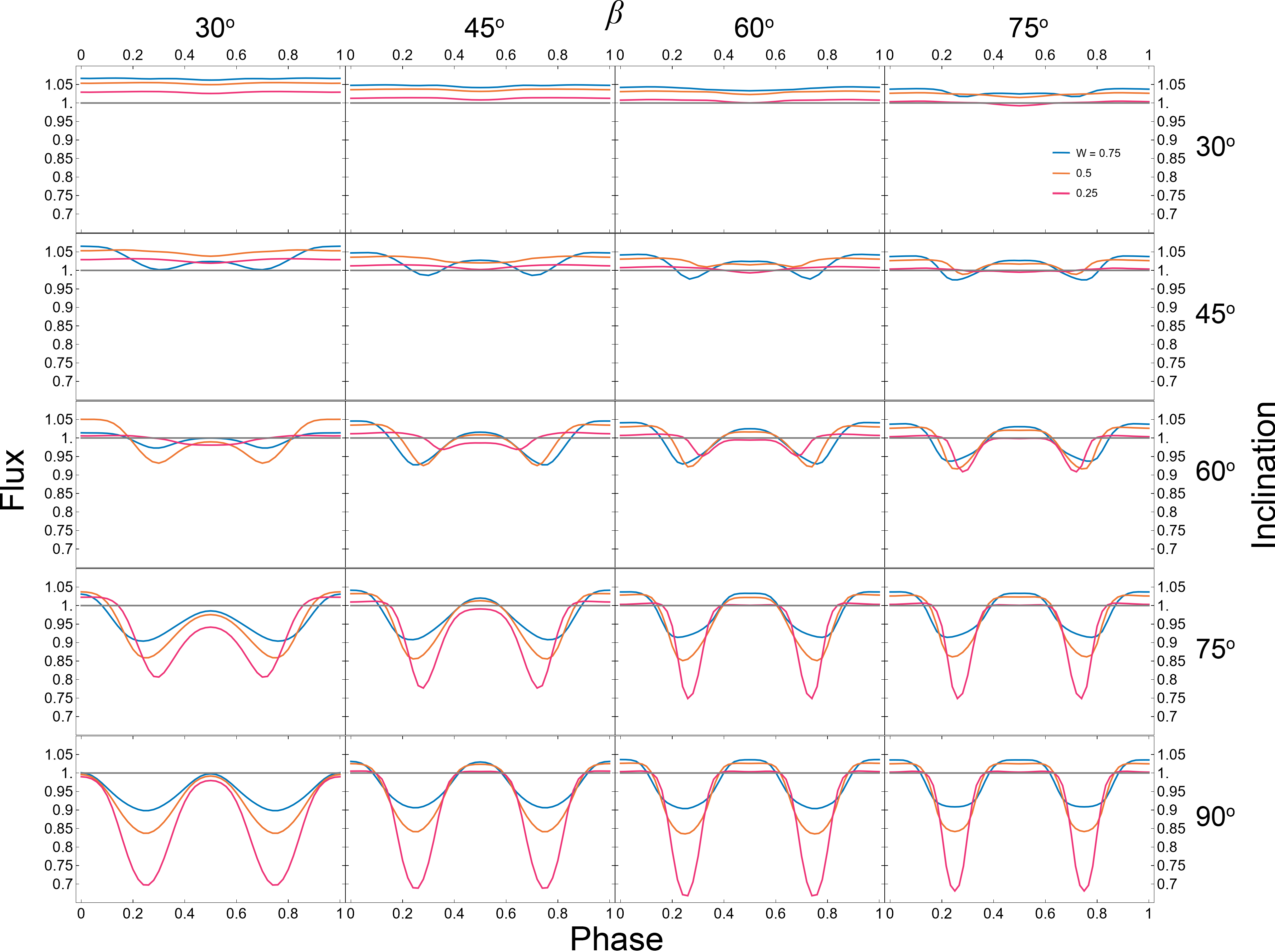}
    \caption{As Figure \ref{lightcurvestau} but with $W = 0.25$, 0.5 and 0.75, for fixed $\tau_{\text{K}} = 2$.}
    \label{lightcurvesW}
\end{figure*}

\begin{figure}
    \centering
    \includegraphics[width = \columnwidth]{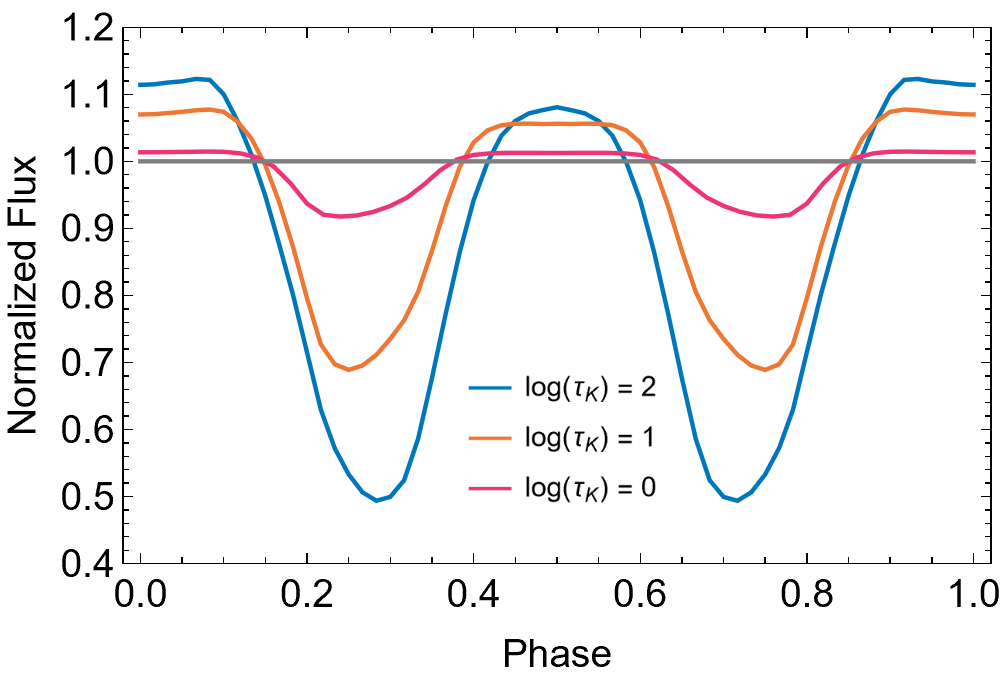}
    \caption{Light Curves at $\beta = i =  75^{\circ}$ and $W = 0.5$ with $\log{(\tau_{\text{K}})} = 0, 1$ and 2. We see that $\tau_{\text{K}}$ has an effect on the duration of emission, as well as the duration of the absorption dips. As the disk becomes significantly optically thick, the magnetosphere will appear slightly more physically thick, meaning the magnetosphere will eclipse the star somewhat sooner in phase when very optically thick.}
    \label{logTau}
\end{figure}

 To investigate the application of scattering radiative transfer further, let us now focus on changing $\beta$. For any $\beta \neq 0$, the 2D axisymmetry present in the field-aligned case disappears, as the plasma trapped in the CM becomes more concentrated at the intersections of the magnetic and rotational equatorial planes (as seen in section \ref{RRMCBO}). Variation in density along the azimuthal coordinate $\phi$ is present, leading to a variation in flux with rotational phase. These intrinsic changes in CM geometry with $\beta$, and the consequent changes in the projected appearance of the CM with $\phi$, are illustrated in Figure \ref{surfaceint}. 
 
 Figures \ref{lightcurvestau} and \ref{lightcurvesW} show flux variation vs. phase while changing $\beta$, varying $\tau_{\rm K}$ for fixed $W$ in Fig.\ \ref{lightcurvestau} and vice versa in Fig.\ \ref{lightcurvesW}. In both cases, at low inclination angles, increasing $\beta$ results in lower emission as the surface area of the CM is intrinsically smaller. Absorption first occurs at $i = 45^{\circ}$, $\beta = 75^{\circ}$ as the CM now  passes in front of the star. $\beta$ has an effect on the duration of absorption, with the lowest $\beta$ having the longest periods of absorption. Likewise, $\beta$ has a significant effect on the nature of the emission. Larger values of beta lead to \say{plateaus} of emission rather than peaks, due to the magnetosphere spending a considerable fraction of a rotational cycle projected off of the limb. Emission is present in all cases except for the $\beta = 30^{\circ}$, $i = 90^{\circ}$ case where the magnetosphere is always eclipsing the star.
 
 Similarly to section \ref{aligned}, variation of the observer's inclination leads to variations in flux. Generally, lower $i$ leads to similar effects seen with the field aligned case, i.e. only emission occurs except for high $\beta$ and low $W$. However, higher $i$ does not explicitly lead to just pure absorption as occurs in section \ref{aligned}. Rather, there are rotational phases at which the magnetosphere can appear entirely off the limb of the star, leading to pure emission.

  For fixed $\epsilon = 0.01$ and $W = 0.5$, Figure \ref{lightcurvestau} shows the variation of flux with rotational phase for 3 selected values of $\tau_{\rm K} = 0.5$, 1 and 2. These values of $\tau_{\rm K}$ correspond to $B_* \sim 5 - 10$ kG (see Figure \ref{tauBVar}). Increasing $\tau_{\text{K}}$ has the effect of increasing the amount of emission when the magnetosphere is projected off the limb of the star, as well as increasing the amount of absorption when eclipses are occurring. This effect is expected via the solution to radiative transfer (equation \ref{eq:Ixy}). For $\tau_{\rm K} \approx 1$, we see emission of $\sim 5\%$. 

  As $\tau_{\text{K}}$ increases, the magnetosphere becomes optically thick at a greater distance from the Kepler radius, which increases the duration of eclipses. However, $\tau_{\text{K}}$ needs to increase an order of magnitude for this effect to become apparent. This is illustrated in Figure \ref{logTau}, which shows the effect of varying the {\em logarithim} of $\tau_{\rm K}$ between  0 and 2 (i.e.\ probing the maximum possible optical depth for the most extreme combination of physically plausible rotation and surface magnetic field strength), for the case of $i = \beta = 75^\circ$.\par
  
 For fixed $\tau_{\rm K}$ and $\epsilon = 0.01$, Figure \ref{lightcurvesW} shows phase variation of flux for 3 selected values of $W = 0.25$, 0.5 and 0.75. This corresponds to a range of possible $B_* \sim 5 - 25$ kG (see Figure \ref{tauBVar}). Similar to Figure \ref{tauandW} we see that flux increases on the order of a few percent with increasing $W$, as well as lower absorption, as less of the star is eclipsed. Varying $W$ also significantly affects the width of absorption dips. Generally, increasing $W$ will make the absorption start earlier and end later in phase. The opposite effect occurs with emission. As the CM spends more time eclipsing the star, it will spend a shorter fraction of a rotational cycle off the limb of the star, which leads to less emission at higher $i$.

\section{Discussion} \label{disc}

\subsection{Comparison to previous results}

Figures \ref{lightcurvestau} and \ref{lightcurvesW} show that electron scattering emission leads to an increase in flux above the continuum with a marginally optically thick magnetosphere (e.g. $\tau_{\rm K} = 1$). However, flux modulation is heavily dependent on $i$. At low inclinations and $\beta$ (e.g. $i \leq 30^{\circ}$, $\beta \leq 45^{\circ}$) flux remains relatively constant, reminiscent to the field aligned case. To get any absorption, a high inclination is needed. The more the observer is looking along the rotational equator, the more the stellar disk will be eclipsed by the magnetosphere, leading to deeper absorption dips. Another key factor that $i$ plays is the shape of the absorption dips. Specifically, for any $i \neq 90^{\circ}$, dips will not be perfectly symmetric about their absolute minimum, due to the observer's line of sight being tilted with respect to the rotational equator. This asymmetry can be seen in similar light curves produced by \cite{2008MNRAS.389..559T}\par

$\beta$ plays a significant role in flux modulation as well, mainly affecting the duration and amount of absorption and emission in phase. In general, at low $i$, the amount of flux above the continuum will decrease with increasing $\beta$. As $i$ increases and absorption starts to take effect, increasing $\beta$ will generally increase the duration of emission, leading to \say{plateaus} of emission, rather than peaks. Likewise there will be less absorption in phases with higher $\beta$, leading to narrower absorption dips in light curves.

The parameter study reflected in Figures \ref{lightcurvestau} and \ref{lightcurvesW} follows the same general strategy used by \cite{2008MNRAS.389..559T}, in which $\beta$, $i$, $W$ and $\tau_{\rm K}$ ($\tau_0$ in \cite{2008MNRAS.389..559T}) were varied. The resulting light curves share many common trends. In general, both  studies show a double-minimum at higher $i$ and $\beta$. Furthermore, \cite{2008MNRAS.389..559T} shows that increasing $W$ increases the duration but decreases the maximum depth of absorption, and that increasing $\tau_{\rm K}$ increases the maximum depth of absorption. 

But the two approaches have key differences, both in the radiative transfer and the underlying RRM model used.
Our inclusion of electron scattering (vs. pure absorption) leads to modest light-curve emission features from scattering by material off  the limb. This also fills in somewhat the net absorption by material seen against the stellar disk, changing the detailed shapes of the light curves, with eclipses that have broader, shallower minima. 

Our CBO-limited density scalings give an overall lower optical depth, with also steeper declines in radius and at azimuths away from the axis of the common equator.
The absorption by these equatorial clouds thus remains in two distinct phases, without the broad merged absorption features seen in \cite{2008MNRAS.389..559T} models when the extended off-equator azimuths are taken to have sufficient optical thickness to occult the star.
Our model predicts at least modest photometric variability for all $(i,\beta)$ combinations except for the special case of an aligned dipole. In contrast, the RRM model used by \cite{2008MNRAS.389..559T} predicted variability only if $i + \beta \gtrapprox 90^\circ$. 

\subsection{Comparison to $\sigma$ Ori E}

While a detailed comparison of observed to synthetic light curves is outside the scope of this paper given that $\sigma$ Ori E's surface magnetic field is not accurately described by the simple tilted dipole model presented here, it is still possible to make an approximate comparison of the broad predictions made by the CBO-modified RRM model.

The light curve of $\sigma$ Ori E shows two absorption dips on the order of $15\%$, which tells us that the inclination must be high, as shown by \cite{2015MNRAS.451.2015O}. For $\tau_{\rm K} \sim 1$ and $W = 0.25-0.5$, we see this level of absorption in our light curves. One emission bump  of $\sim 4\%$ above the continuum at phase 0.6 is seen in the light curve of $\sigma$ Ori E. Our light curves show that this level of emission is possible at high inclinations with optical depths of order unity and $W \sim 0.5$. 

Using equations from section \ref{escattering}, we can calculate a specific value of $\tau_{\rm K}$ for $\sigma$ Ori E. We start with the stellar, magnetic, and rotational parameters determined by \cite{2019MNRAS.490..274S}, who derived $M_* = 7.9~{\rm M_{\sun}}$, $R_* = 3.39~{\rm R_{\sun}}$, $B_* = 10$ kG, and $R_{\rm K} = 2.69~R_*$ by requiring the fundamental parameters, projected rotation velocity, etc.\ be consistent with the large inclination determined by \cite{2015MNRAS.451.2015O} via ARRM modelling of the light curve. These parameters lead to a $\tau_{\rm K} \approx 0.21$.\par


Looking at the light curves  for $\tau_{\rm K} = 0.5$ and $W = 0.5$, we see that the amount of absorption and emission is not nearly enough to account for the levels seen in the light curve of $\sigma$ Ori E; obviously, even weaker emission, and shallower absorption, would be seen using $\tau_{\rm K} = 0.21$ and $W = 0.2$. 

One possible resolution of that may be that the Kepler radius inferred by \cite{2019MNRAS.490..274S} is too large. Referring to Fig.\ \ref{taukVar}, an optical depth of order unity for a star with $\sigma$ Ori E's surface polar magnetic field strength would require $R_{\rm K} \sim 1.9~R_*$. This would imply $R_* \sim 5 {\rm R_\odot}$, in turn necessitating $i \sim 40^\circ$. Such a small $i$ was clearly ruled out by \cite{2015MNRAS.451.2015O}, who found by contrast that $i > 80^\circ$ with high confidence. Furthermore, $R_{\rm K} \sim 1.9 R_*$ is inconsistent with the star's H$\alpha$ emission, which peaks at a velocity corresponding to $3 R_*$ \citep{
2020MNRAS.499.5379S}.

Examining the light curves generated in the present work, one that stands out as a potentially appropriate fit for $\sigma$ Ori E is that with $i = 75^{\circ}$, $\beta = 30^{\circ}$, $\tau_{\rm K} = 2$ and $W = 0.25$. First, this light curve only has one phase window of emission present, as is the case with $\sigma$ Ori E. It also produces an appropriate amount of emission, but only slightly too much absorption, which tells us that $\tau_{\rm K}$ should be slightly less than 2. This is a good match both to the inclination determined by \cite{2015MNRAS.451.2015O}, $i = 85^\circ$, and to the $\beta = 38^\circ$ and $W = 0.22$ determined by \cite{2019MNRAS.490..274S}. 

One main difference between our light curves and the observed light curve of $\sigma$ Ori E is the timing of the phases. For $\sigma$ Ori E, emission occurs at phase $\sim 0.6$, whereas emission occurs at phase $\sim 0$ in our models. This difference is based on where phase 0 is defined. The $\sigma$ Ori E light curve has phase 0 at the point of maximum absorption. In our models, phase = 0 is chosen to be when the magnetic field axis is pointing towards the observer. Furthermore, $\sigma$ Ori E shows a secondary minimum about 0.2 cycles before emission begins. Indeed this phase difference is seen between the second minimum at phase 0.75 and the onset of emission at phase 0.95 in the $\beta = 30^{\circ}, i = 75^{\circ} $ and $W = 0.25$ light curve.

It therefore seems that the angular and rotational parameters of the model most resembling the observed light curve of $\sigma$ Ori E are well-matched by the values independently determined via other means. This then leaves the question of the low value of $\tau_{\rm K} \sim 0.2$ predicted for the star. One possibility is that the correction factor, taken to be $c_{\rm f} = 0.3$ based on calibration by 2D MHD simulations for the field-aligned ($\beta = 0$) case \citep{2020MNRAS.499.5366O}, is simply inappropriate for models with non-zero tilt. We find that in order to obtain an $\tau_{\rm K} \sim 1$ for the provided $\sigma$ Ori E parameters $c_{\rm f}$ needs to be of order unity.\par

\subsection{Application to other $\sigma$  Ori E variable stars}

While the development of the model presented here was motivated by the need to explain the otherwise inexplicable photometric emission bump in $\sigma$ Ori E's light curve \citep[][]{2015MNRAS.451.2015O}, there are a number of other stars exhibiting $\sigma$ Ori E variability consistent with an origin in centrifugal magnetospheres \citep[see][for the most recent catalogues of these populations]{2013MNRAS.429..398P,2020MNRAS.499.5379S}. In most cases, $\mu$mag-precision light curves from the {\em Kepler} or Transiting Exoplanet Survey Satellites ({\em TESS}) space telescopes are now available \citep{2010Sci...327..977B,2014PASP..126..398H,2015JATIS...1a4003R}. 

A detailed analysis of the results from space photometry is outside the scope of this paper and will require a dedicated work. However, a preliminary analysis of the {\em TESS} light curves of a number of magnetic chemically peculiar stars by \cite{2020pase.conf...46M} has demonstrated that several objects (in particular, Landstreet's Star HD\,37776, as well as HD\,64740) exhibit `warped' light curves characterised by the presence of a large number of harmonics of the rotational frequency. \citeauthor{2020pase.conf...46M} were unable to reproduce these highly structured light curves with photospheric chemical spot models, which should in general produce light curves with at most one or two harmonics of the rotation frequency; since these stars generally display H$\alpha$ emission, they inferred that the large number of harmonics are a consequence of circumstellar material. 

In the cases of HD\,37776 and HD\,64740, it is an open question whether their structured light curves are a result of eclipses or electron scattering emission. Both stars have very weak H$\alpha$ emission, and both stars furthermore demonstrate eclipses in H$\alpha$ \citep{2020MNRAS.499.5379S}. Furthermore, HD\,37776 has an extremely complex, multipolar magnetic topology \citep{2011ApJ...726...24K}, which makes application of the dipolar models developed here fundamentally inappropriate. An additional complication is that the light curves of both stars very obviously contain prominent contributions from their photospheric abundance spots. Disentangling the photospheric and circumstellar contributions would require Doppler mapping of the abundance distributions and forward modelling of the contribution of spots to the light curve \citep[as was done for $\sigma$ Ori E by][]{2015MNRAS.451.2015O}. 

An answer to the question of whether other stars might show photometric brightening due to electron scattering in their CMs might be found in the case of HD\,37017. This star's geometry \citep[$i=38^\circ$, $\beta=56^\circ$;][]{2019MNRAS.490..274S} is not expected to produce eclipses, and indeed it shows no sign of eclipses in H$\alpha$, with a prominent emission profile indicative of a large CM \citep{2020MNRAS.499.5379S}. Its {\em TESS} light curve was presented by \cite{2020svos.conf..177S}, who noted that it displays the same structure inferred to be circumstellar by \cite{2020pase.conf...46M}. Furthermore, the structure is distributed across the light curve, rather than being concentrated near magnetic null phases at which eclipses are expected. HD\,37017 is therefore an excellent candidate for photometric emission from an electron scattering CM. 

Another star which shows signs of such emission is HD\,144941, a magnetic extreme Helium star with a highly structured light curve, H$\alpha$ emission, and a magnetic geometry that is probably not consistent with eclipses \citep{2018MNRAS.475L.122J,2021MNRAS.tmp.1941S}.

While not remarked upon by the authors, visual inspection of the the {\em MOST} light curve of the extremely rapidly rotating star HR\,5907 \citep[$P_{\rm rot} \sim 0.5$~d;][]{2012MNRAS.419.1610G} is suggestive of structure. This star demonstrates an H$\alpha$ eclipse near phase 0, however the light curve structure is apparent at other phases as well. Since HR\,5907 has extremely strong H$\alpha$ emission comparable to that of $\sigma$ Ori E \citep{2020MNRAS.499.5379S}, it is an excellent candidate for electron scattering emission. The comparably rapidly rotating star HR\,7355 \citep{2010MNRAS.405L..51O,2010MNRAS.405L..46R,2013MNRAS.429..177R} is also an excellent candidate for photometric emission, given both its strong H$\alpha$ emission and the prominent eclipses in its {\em Hipparcos} light curve \citep{2008A&A...482..255R}. 

A feature of the present models that does not seem to be reproduced in the published light curves of the few stars in which photometric emission can be reasonably inferred are the broad, flat maxima, which stand in contrast to the relatively sharp bumps prominently seen in the case of $\sigma$ Ori E. This is most likely an inevitable consequence of the simple tilted dipole geometry adopted here, which results in a plasma distribution that is symmetrical about the magnetic axis. This geometry is certainly not applicable to $\sigma$ Ori E itself, which has prominent contributions from higher-order multipoles to its surface magnetic field \citep{2015MNRAS.451.2015O}. The same is likely true of other $\sigma$ Ori E variable stars, which frequently possess longitudinal magnetic field curves indicative of departures from a purely dipolar topology \citep[e.g.][]{2018MNRAS.475.5144S}, as well as asymmetries in their H$\alpha$ emission profiles that likely arise from these deviations from a dipole \citep[e.g.][]{2020MNRAS.499.5379S,2021MNRAS.504.3203S}. This indicates that modelling the light curves of these stars will probably require not just inclusion of the photospheric contribution via Doppler mapping of chemical spots, but Zeeman Doppler imaging of the surface magnetic field in conjunction with potential field extrapolation using an arbitrary RRM model, i.e.\ the same process that was performed by \cite{2015MNRAS.451.2015O} for $\sigma$ Ori E. Such models are not currently available in most cases, but can probably be determined using high-resolution spectropolarimetric data already in hand thanks to large surveys such as the Magnetism in Massive Stars \citep[MiMeS;][]{2016MNRAS.456....2W} collaboration. While such an effort would be significiant, it would provide unique information on the density structure inside the magnetospheres of rapidly rotating hot stars.

\section{Summary and Future Work} \label{sum}

We have modified the RRM model to utilize a density scaling based upon CBO, which predicts higher and more centrally concentrated density distributions sufficient to produce photometric emission via electron scattering. Using the RRM-CBO model, we have presented light curves accounting for both photometric absorption and emission, as a means of testing the hypothesis that an emission bump in the light curve of the B2\,Vp star $\sigma$ Ori E may originate due to scattering in its circumstellar magnetically confined plasma. With optical depth of order unity, we can generate light curves that display emission and absorption of the same magnitude as that seen in $\sigma$ Ori E's light curve, with the best-matching light curve having similar angular and rotational parameters to those independently derived via other means.

We have conducted a parameter study examining the effects of varying the inclination angle, the magnetic obliquity angle, the optical depth at the Kepler corotation radius, and the critical rotation parameter; all parameters affect the amount of absorption and emission. 

While the present work focused on light curves in the continuum using a centered dipole of first order as the magnetic field, future work can include,

\begin{itemize}
    \item Stars have more complex fields than a dipole of first order. Similar light curves to the ones in the present paper should be created using more complex fields.
    \item Scattering in the continuum is only one form of emission that can be modeled. Other forms of emission can and should be modeled, namely H$\alpha$ line profiles.
    \item Magnetohydrodynamic simulations are another way to model magnetospheres, and show phenomena not displayed by the RRM model. The RRM model used in the present work used scalings inspired by recent MHD simulations. Photometric light curves should be created using MHD for further direct comparison to RRM and MHD model predictions.
    \item $\sigma$ Ori E inspired the work in this paper, however, we did not directly compare our generated light curves to the photometric light curve of $\sigma$ Ori E. Direct comparison between theoretical light curves and real light curves will need to be done in the future.
    \item Light curves of massive stars have been fit with absorption and elemental surface abundances. Light curves should now be fit with absorption, surface abundances, and continuum scattering.
\end{itemize}

A central consequence of this work, in contrast to previous RRM models that only considered eclipses, is that photometric variation should occur for essentially all combinations of angular parameters. Overall we find that electron scattering yields a small but significant amount of emission so long as the optical depth is of order unity, at a level which can be easily probed using modern $\mu$mag space photometry. 

\section*{Acknowledgements}
IDB acknowledges student research support from the Department of Physics and Astronomy and the Bartol Research Institute at the University of Delaware. AuD acknowledges support by the National Aeronautics and Space
Administration through Chandra Award Number TM1-22001B issued by the
Chandra X-ray Center which is operated by the Smithsonian Astrophysical
Observatory for and on behalf of NASA under contract NAS8-03060. AuD
also acknowledges support from Pennsylvania State University
Commonwealth Campuses Research Collaboration Development Program. Data
was generated through this support from the Institute of Computational
and Data Sciences. MES acknowledges the financial support provided by the Annie Jump Cannon Fellowship, supported by the University of Delaware and endowed by the Mount Cuba Astronomical Observatory. 

\section*{Data Availability Statement}
The software presented in the current work is available upon request from the authors.

\bibliographystyle{mnras}
\bibliography{export-bibtex.bib}

\end{document}